\documentclass[preprint]{aastex}

\shorttitle{Spectral Space of QSOs}
\shortauthors{Boroson \& Lauer}

\begin{document}

\title{Exploring the Spectral Space of Low Redshift QSOs}

\author{Todd A. Boroson and Tod R. Lauer}
\affil{National Optical Astronomy Observatory, Tucson, AZ 85719}
\email{tyb, tlauer@noao.edu}

\begin{abstract}
The Karhunen-Lo\`eve (KL) transform can compactly represent the information 
contained in large, complex datasets,  cleanly eliminating noise from the data
and identifying elements of the dataset with extreme or inconsistent characteristics.
We develop techniques to apply the KL transform to the 
4000-5700 \AA\  region of 9,800 QSO spectra with $z < 0.619$ from the SDSS archive.
Up to 200 eigenspectra are needed to fully reconstruct the spectra in this sample 
to the limit of their signal/noise.
We propose a simple formula for selecting the optimum 
number of eigenspectra to use to reconstruct any given spectrum,
based on the signal/noise of the spectrum,
but validated by formal cross-validation tests.
We show that such reconstructions can boost the effective signal/noise
of the observations by a factor of 6 as well as fill in gaps in the data.
The improved signal/noise of the resulting set will allow for better measurement
and analysis of these spectra.
The distribution of the QSO spectra within the eigenspace
identifies regions of enhanced density of interesting subclasses,
such as Narrow Line Seyfert 1s (NLS1s).
The weightings, as well as the inability of the eigenspectra to fit some of the 
objects, also identifies ``outliers,'' which may be objects that are not valid members of 
the sample or objects with rare or unique properties.  We identify 48 spectra from the 
sample that show no broad emission lines, 21 objects with unusual [O III] emission line 
properties, and 9 objects with peculiar H$\beta$ emission line profiles.  We also use 
this technique to identify a binary supermassive black hole candidate.
We provide the eigenspectra and the reconstructed spectra of the QSO sample.
\end{abstract}


\section{A Tool to Represent and Visualize the Information in Large Datasets}

The advent of large samples of QSO spectra, particularly that from the Sloan Digital 
Sky Survey (SDSS) \citep{yor00}, has allowed richer analyses of their properties than has been 
possible previously.  At the same time, these datasets are 
too large for manual inspection and analysis of individual observations over the 
entire sample; thus, new tools are needed to understand their content.  The 
information sought is sometimes related to common features and correlations 
among the characteristics of the sample, and it is sometimes related to the most 
extreme exceptions or outliers from these correlations.  In either case, understanding 
how to reduce the dimensionality of the data can be helpful.  We advance the method 
of the Karhunen-Lo\`eve (KL) transform (also called Principal Components Analysis) 
as a powerful solution to this problem.

The spectra of QSOs are good examples of where this approach is needed, and 
indeed KL transforms have already played an important role in their analysis.  Although 
the spectra are dominated by common features, broad emission lines from abundant 
elements, narrow forbidden emission lines, resonance line absorption from intervening 
and associated material, the details of the physical processes at work are obscured by 
a complicated and unknown interplay among many parameters.  No simplifying tool 
like the HR diagram has emerged for QSOs.  

The KL transform creates a new set of orthogonal basis vectors, or eigenvectors, for 
a sample, ordered in decreasing importance in accounting for the sample variance.  
In the case of a sample that consists of spectra, these eigenvectors can be represented as
{\it eigenspectra,} which convey a spectral view of the relationships inherent in the 
eigenvectors. The eigenvalue that corresponds to each eigenvector is a measure of 
the relative importance of that eigenvector in accounting for the variance within the 
sample.  Highest weight goes to the features with the largest systematic 
variation within the sample, and lower weight goes to the features in which noise 
or a unique spectral characteristic dominates.  The commonality of features in QSO 
spectra indicate that their 
complexity is of much lower dimensionality than the number of spectral bins or the 
number of objects in a large sample.  The KL transform can take advantage of that 
fact to concentrate the information in the dataset, which allows the de-noising of the 
data, as well as the identification of outliers.

For samples that are as large as several thousand spectra, it is of interest to quantify 
the variation within the entire ensemble to better understand the physical 
parameters that drive the variation. This type of analysis has been applied to QSO 
spectra by several researchers, including \citet{fra92}, \citet{sha03}, and \citet{yip04}.  
These studies have emphasized the use of the eigenvectors to develop a physical 
understanding of the properties of the sample. However, the KL transform 
is a purely mathematical technique, thus efforts 
to attribute physical meaning directly to the basis vectors have met with limited success.

We thus ignore the issue of what any particular 
eigenspectrum ``means,'' instead choosing to exploit the mathematics of the KL transform 
and the ensemble of the eigenspectra as a means to explore the 
sample as well as to represent the spectra of individual QSOs.  This leads to the 
twin goals of the present paper. First, because true sample-wide information is found 
in the most important (lowest) eigenvectors, and the object-specific noise is relegated 
to the least important (highest) ones, the set of eigenvectors can be used to  
reconstruct the spectra without including much of the noise. That is, rather than 
processing the spectra with any number of general purpose linear filters to reduce 
noise, we will use the eigenspectra to isolate and preserve the information content 
in any given QSO.

A second goal is to detect outliers, that is, objects that are not easily or well 
described by the dominant eigenvectors of the sample. This allows the establishment 
of a sample with well defined properties and the identification of objects that might 
have been mistakenly included, or might be particularly interesting due to rare or 
unusual properties. As we will show, there is some tension between the goals of 
providing a reduced noise representation of any individual QSO, while detecting 
other individual QSOs having properties that are rare within the sample.  

The technique of the KL transform is well established.  This paper demonstrates a
new application of this technique to a particular dataset.  This sample was chosen 
because of 
the richness of the region of the spectrum observed for these objects, and because 
much study has already gone into the properties of the lines and continuum in this
region.  Section 2 
describes the real application of the KL transform to the data and demonstrates the 
validity and power of the technique.  Section 3 presents statistical measurements of 
the properties of the sample in the context of the eigenspace.  Section 4 presents the 
outlier objects and discusses their properties.  The conclusions are recounted in 
Section 5.

\section{The Application of the KL Transform to the SDSS QSOs}

In practice, the application of the KL transform to a real data set requires adoption 
of a number of assumptions and techniques due to the deficiencies and quirks of 
real data. An excellent description of these issues is presented by \citet{con99}, 
who emphasize the use of the KL transform for filling in gaps or improving 
signal-to-noise in individual objects within a sample.

\subsection{The QSO Sample}

The QSO sample is a low redshift subset from
the fourth version of the SDSS Quasar Catalog \citep[SDSSQ4]{sch07}.  This catalog 
is based on the fifth SDSS data release (DR5), and is limited to objects that have 
absolute I-band magnitude brighter than $-22$, and have at least one emission line 
with a full width at half-maximum (FWHM) larger that 1000 km s$^{-1}$.  The QSOs 
having redshifts below 0.619 (9,800 objects) were drawn from this catalog.  This 
upper redshift cutoff allows complete coverage of the 4000 - 5700 \AA\ rest frame 
region.

The SDSS spectra cover the observed wavelength region 3800 - 9200 \AA\ at a spectral 
resolution of R $\simeq$ 2000.  Each reduced spectrum includes a calibrated flux vector, 
a noise vector, and a mask vector, the last of which indicates the regions in which 
data are missing or suspect due to processing problems or other effects. Derived quantities, 
including the galactic absorption in the SDSS photometric bands and the adopted 
redshift, are given in the FITS header. The sample is not statistically 
complete; some objects are selected as QSO candidates on the basis of broad-band 
colors, while others are chosen on the basis of proximity on the sky to radio or x-ray 
sources.

Each spectrum was corrected for galactic absorption and rebinned to a common rest 
wavelength scale, covering the range 4000 - 5700 \AA\ . The redshifts adopted are 
those from the SDSSQ4 catalog, which are based primarily on the SDSS {\it spectro1d} 
pipeline.  These are derived by automatic identification of emission lines, which are 
constrained to have certain relative positions \citep{sto02}.  The resulting redshifts 
represent a weighted average, and account, in a limited way, for the velocity offset 
seen in some emission lines in some QSOs \citep{gas82, ric02}.  Our rebinning 
used a logarithmic dispersion of $d\lambda/\lambda = 10^{-4},$ resulting in 1540 
pixels per spectrum.  

We computed an average signal-to-noise ratio for 
each rebinned spectrum by averaging the ratio of each unmasked flux pixel to its corresponding 
noise pixel within the wavelength range of interest.  These ratios range from 1.3 to 95, 
but 90\% have a ratio larger than 6, and half have a ratio larger than 12. 

\subsection{Derivation of the Eigenspectra}

We construct the eigenspectra from a subset of the sample comprising those QSOs with (a) 
average signal-to-noise ratio above 25 in the 
4000 - 5700 \AA\ region, and (b) no more than 100 pixels flagged as bad.  This latter 
criterion prevents the possibility of the red end of the spectra of the higher redshift objects 
contributing excess noise or errors from poor subtraction of the OH bands that dominate
the sky emission.  These restrictions produced a sample of 1039 QSOs.  
These are distributed approximately uniformly over the luminosity range spanned
by the entire sample, $-26.5 < M_i < -22.0$. 

The process of producing the basis vectors includes several steps.  Each spectrum 
is normalized by its scalar product \citep{con95}, and a 1039 $\times$ 1039 cross correlation 
matrix, C, is calculated, representing the cross product of each spectrum with all the others.
Singular value decomposition is then used to find the matrix, U, for which

\begin{equation}
U^{T}CU=\Lambda ,
\end{equation} where $\Lambda$ is a diagonal matrix, whose diagonal elements are the eigenvalues.  
The matrix $U$ contains the eigenvectors as columns. These eigenvectors are 
multiplied by the input spectra to produce the eigenspectra of the system.
We found that the first 200 eigenspectra were sufficient to represent the sample.

In practice, the presence of bad data or data gaps in the set of 1039 high S/N QSOs 
used to specify the eigenspectra required
the derivation of the eigenspectra to be done iteratively.
During the initial calculation of the eigenspectra,
pixels that were flagged as bad in the mask array were not included in the calculation
of the cross-correlation products.
After the initial calculation of the eigenspectra was completed,
the bad or missing data in any given input QSO spectrum were replaced with 
reconstructed data derived from the eigenspectra-based representation of the spectrum.
The corrected spectra were then used to produce a new revised set of eigenspectra,
which in turn were used to produce better reconstructions of the deleted data.
This sequence was repeated until the corrections converged.

The procedure for determining how a given QSO spectrum is represented by the eigenspectra 
should account for both the uncertainty in each pixel and the gaps or flagged bad 
regions in the spectrum.  After normalization, a spectrum is projected onto the 
eigenspectra, using, for each pixel, a weight corresponding to 1/$\sigma^2$, where 
$\sigma$ is the element from the noise vector for that spectrum.  Gaps (pixels masked
as bad) are not included.  
The inclusion of the weights and gaps result in the basis spectra not being orthogonal, 
and so a correction to these projections must be determined.  This correction comes 
from the elements of the matrix that results by inverting the cross correlation matrix of 
the eigenspectra, computed with the same weights and gaps \citep[equation 5]{con99}. 
These corrected projections, $a_i$, allow the reconstruction of the given spectrum with 
some chosen number of the eigenspectra.  This reconstruction can be shown to 
minimize the $\chi^2$ statistic:

\begin{equation}
\chi^2 = \sum_\lambda(f_\lambda-\sum_ia_ie_{i\lambda})^2/\sigma_\lambda^2 ,
\label{eqn:recon}
\end{equation}
where $f_\lambda$ is the input spectrum and $e_{i\lambda}$ is the $i^{th}$ eigenspectrum.  
In practice, we used $\chi^2/\nu$, the reduced $\chi^2$, where $\nu$ is the number of 
spectral pixels minus the number of eigenvectors used. 

Ultimately, we are interested in producing a coherent sample that is well described by 
the optimum number of eigenspectra.  It is thus necessary to make certain that 
objects that are not proper members of this sample do not contribute to 
the eigenspectra.  Each input spectrum was reconstructed using an optimum number of 
eigenspectra derived from our analysis below.  
Objects that had a resulting $\chi^2/\nu$ value larger than 1.5 were inspected and 
compared with the reconstruction.  Seven of these objects were removed from the 
sample used to compute the eigenvectors (though they were retained in the overall 
sample and will be discussed below as outliers).  Then, the process was then repeated 
with the remaining 1032 objects, producing the final set of eigenspectra.  

\begin{figure}
\epsscale{.90}
\plotone{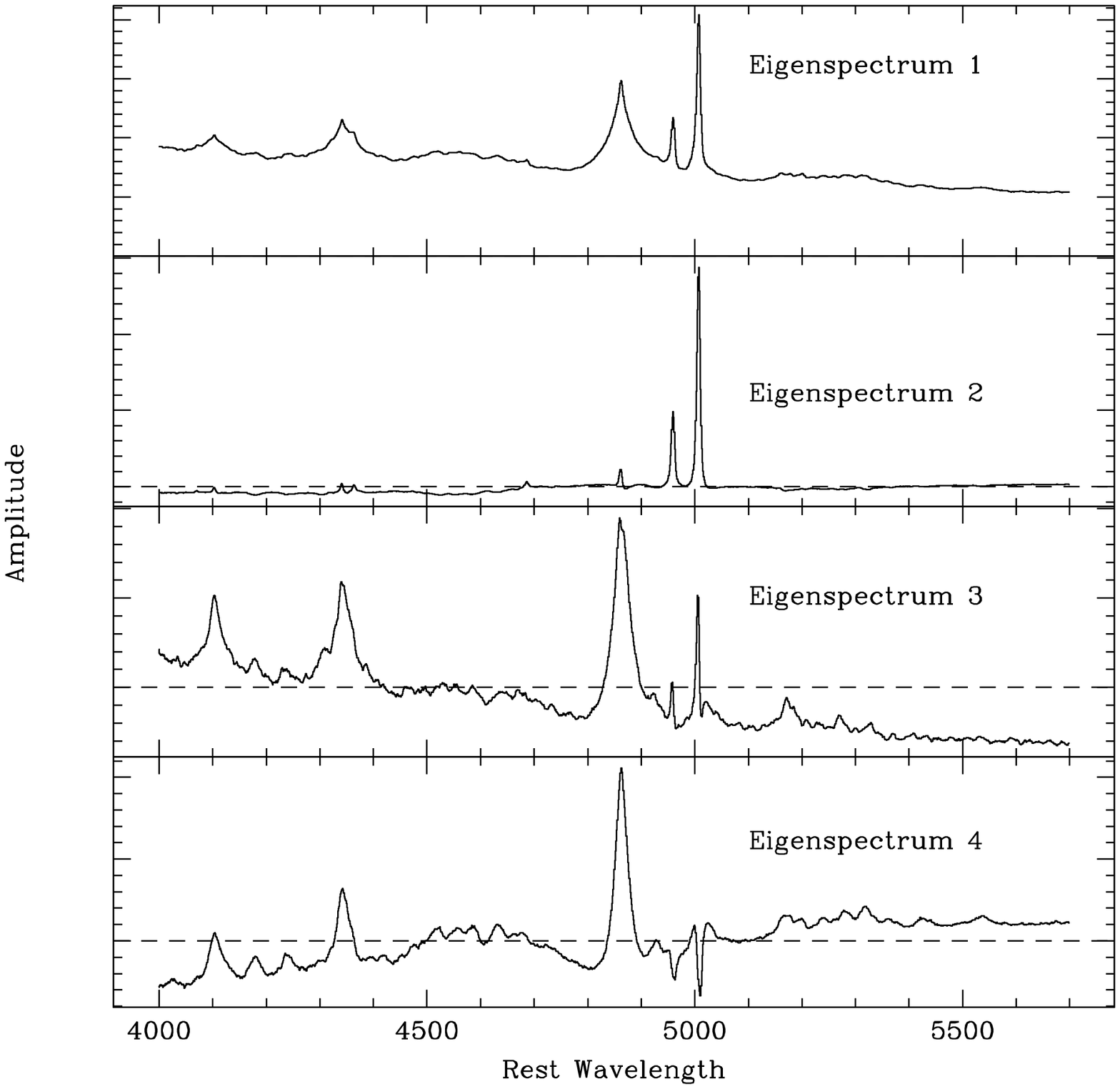}

\caption{First four eigenspectra.  Note that the vertical scale of each panel is different.  
Each panel shows the zero level with a dashed line.  For eigenspectrum 1, the zero level 
is at the bottom of its panel.  The other eigenspectra all have mean values of zero.}
\label{fig:eig}
\end{figure}

Although it is not our goal to interpret the eigenspectra in a physical way, it is helpful to 
recognize some of their properties.  Figure \ref{fig:eig} shows the first four eigenspectra (which 
we term E1, E2, etc.).  Note that the first 
eigenspectrum is the mean of all the normalized input spectra.  It is the only eigenspectrum 
with a mean value that differs significantly from zero. We have chosen to retain the 
mean as the first eigenspectrum, but its unique meaning will be important. The second 
eigenspectrum shows features reminiscent of the \citet{bor92} eigenvector-1 correlations; 
the narrow line components are seen in a positive sense, while the Fe II emission is seen 
as negative.

\begin{figure}
\epsscale{.90}
\plotone{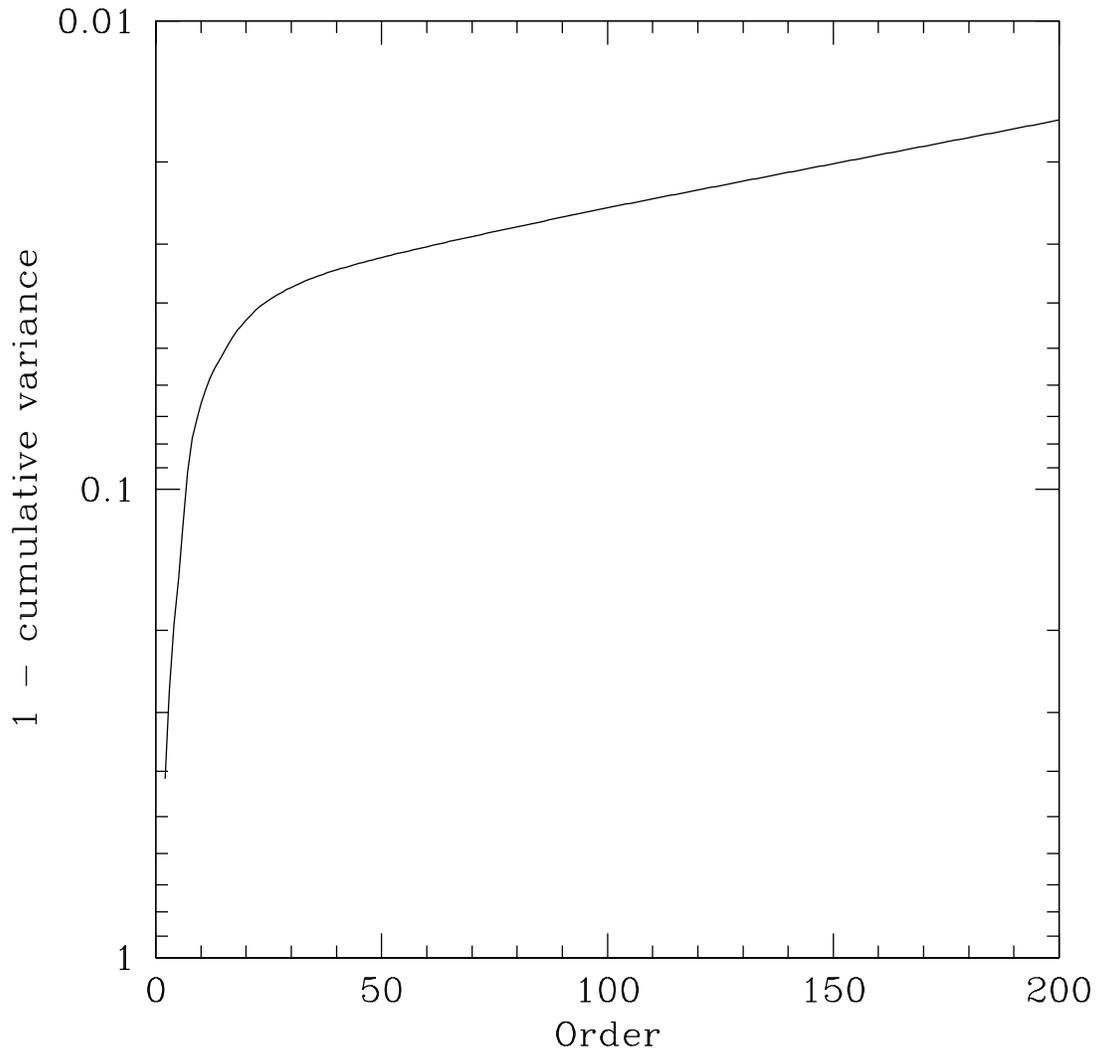}

\caption{This plot shows how cumulative sample variance varies as a function of
eigenspectra number (excluding the first eigenspectrum, which is the mean QSO spectrum for
the sample).  The total variance is normalized to unity, and the ordinate actually
plots the residual variance not explained by the first $N$ eigenvectors, so that
the effects of the higher eigenvectors can be more easily visualized.  Note
that most of the sample variance is explained by the first $\sim30$ eigenspectra.}
\label{fig:var}
\end{figure}

The eigenvalues represent the amount of variance accounted for by each 
corresponding eigenvector.
The first eigenvalue is very large, but it refers to the fraction of 
variance represented by the mean spectrum.  After 
removing this from the sum, Figure \ref{fig:var} shows the cumulative fraction of variance 
compared to the mean spectrum that the inclusion of each successive eigenvector 
accounts for.
Clearly the great bulk of the
variance is captured by the first $\sim30$ eigenspectra.
The first 200 eigenspectra are available in tabular form electronically.  
Table \ref{tab:es} describes the content and format of each column.

\subsection{Using the Eigenspectra to Reconstruct the QSO Spectra}

One major goal is to use the eigenspectra to recover the best estimate
of the underlying spectrum of a given individual QSO in the presence of noise,
using the reconstruction algorithm presented in eq.\ (\ref{eqn:recon})
\citep[also see the discussion in][]{con99}.  A key part of the problem
is developing objective criteria to estimate how many eigenspectra, $N_{opt},$ a will be
used to represent the selected QSO.  However, we also want to
use the reconstructions to test whether or not the QSO might be an ``outlier,"
or an unusual member of the sample, the second major goal of our analysis.
Ignoring this problem could drive us to make $N_{opt}$
large enough to fit rare features in the spectrum,
thus potentially suppressing our recognition of them.  In contrast, detection of outliers
means basing the set-size on generic properties, such as S/N of the QSO spectrum,
rather than its specific goodness of fit.

Suppressing noise in the eigenspectra representation of a QSO means limiting the reconstruction to 
use only the lowest (most fundamental) eigenspectra so that the resulting spectra include 
as much of the ``sample signal,'' but as little of the noise as possible. The derivation of 
the eigenbasis vectors can produce as many eigenspectra as there are input spectra. If 
all eigenspectra are used to reconstruct the input spectra, then this reconstruction will 
result in spectra that are identical to the input spectra. The noise is unique to a given 
input spectrum so its contribution is relegated to the highest eigenspectra.  However, 
real features that may be unique to a given object --- or a small subset of the sample --- will 
also be found only in the highest eigenspectra.  It is important to remember that the set of 
spectra do not represent equivalent observations
of objects from the same parent distribution.  There are clearly objects with extreme properties,
and the spectra span a considerable range of signal-to-noise ratio.  Thus, we cannot adopt
a single $N_{opt}$ to use for the entire sample.

\subsubsection{Basing Reconstruction on the ``Estimated Risk'' Criterion}

In general, all reconstructions of the individual QSO spectra always use the
eigenspectra in the order of their significance
(as determined from the high S/N subsample),
with their specific weights determined from eq. (\ref{eqn:recon}) for the QSO in question.
The issue discussed in this section is how to determine how far down the
list of eigenspectra one should go to reconstruct any given QSO.

Our initial work focussed on use of the traditional $\chi^2/\nu$ statistic.
The $\chi^2/\nu$ value, itself, does not reach a minimum as we add more eigenspectra,
but will continue to decrease to zero as more and more noise is fitted.
One obvious metric would be to increase $N_{opt}$
until the reduced $\chi^2$ for the reconstruction of a given spectrum goes below unity.
While this might produce the most justifiable result for a single spectrum considered
in isolation, it has three drawbacks when the sample is considered as a whole.
First, the expected distribution of reduced $\chi^2$ values is not a delta function at unity,
but is a distribution with some width.
Insisting that all reconstructions proceed to this point will underfit
some spectra and overfit others. Second, it presumes that the noise values are accurate, since it is
when the difference between the model and the data are equivalent to the noise that the
reconstruction is stopped. Third, and most important, it allows the reconstruction to fit
rare features that might qualify a spectrum as an outlier, thus compromising the second
goal of our analysis.

Our derivation of $N_{opt}$ was provided instead by using
``estimated risk,'' a statistic that is similar to 
$\chi^2/\nu$ but does reach a minimum for a specific number of eigenspectra.
The risk is calculated using the technique of
cross-validation \citep{wbook,htfbook}.  This procedure involves excluding a subset of the data from
the fit, and then evaluating the difference between the data and the model for only the
pixels that have been excluded from fitting.  After doing this multiple times, excluding a different
part of the data each time, the estimated risk or the cross-validation value, 
which measures the success of the fit for all data, can be calculated.
The risk will be large both when too few eigenspectra are used and the model
departs systematically from the data (too much bias admitted),
as well as when too many eigenspectra are used and noise has been added
(too much variance admitted).  In essence, for a given $N_{opt}$
the test shows how well the KL transform specified by the data remaining in a given
QSO spectrum can represent the missing data.  If too few eigenspectra are used, then
adding in more provides a better match to the missing data, and the risk decreases.
A point is reached, however, after which increasing the number of eigenspectra
causes the KL transform to start fitting the noise in the preserved portion
of the spectra, thus actually degrading the representation of the missing data,
and hence increasing the risk.

For each QSO in the sample, we performed a 5-fold cross-validation. 
In detail, this entailed performing a large ensemble of trial reconstructions
for each spectrum.
For each trial, 20\%\ of the pixels were excluded from the spectrum,
a pre-selected number of eigenspectra were fitted to the remaining data,
and the contribution of the excluded pixels to the risk value was calculated.
Subsequent trials were conducted with a different 20\%\ of the pixels excluded,
holding the number of eigenspectra fixed, until all the pixels
had been excluded once.
The five values were then combined to produce the estimated risk for
the reconstruction of that particular spectrum with that particular number of eigenspectra. 
This procedure was repeated five times and the results averaged.
The trials for each spectrum were then
repeated, as the number of eigenspectra was increased up to 
200 in order to find number producing the minimum risk value.
For the high signal-to-noise spectra that had been used in the determination of the
eigenspectra, we used alternate sets of eigenspectra,
produced without the particular spectrum that we were fitting.  

\begin{figure}
\epsscale{1.}
\plotone{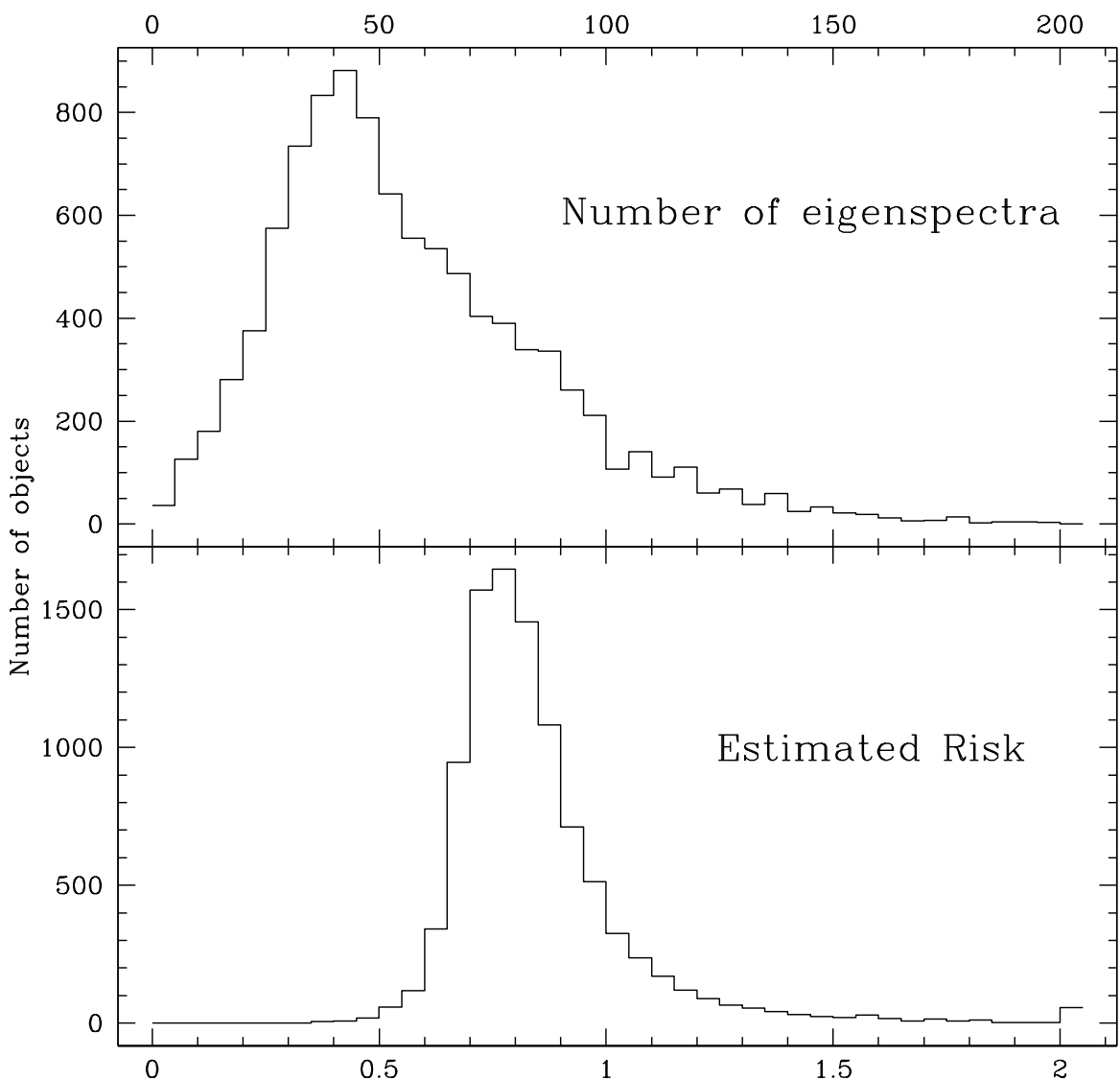}

\caption{The top panel shows the Distribution of the optimal number of eigenvectors
determined to minimize the estimated risk 
(through 5-fold cross-validation) for the
9800 individual spectra.
The bottom panel shows the
resulting distribution of risk values.  The risk values in this context may be regarded
as a normalized $\chi^2$ measure.  Note the the mean risk and its distribution
resembles a $\chi^2$ distribution.}
\label{fig:opt}
\end{figure}

Figure \ref{fig:opt} shows the resulting distribution of optimum number of eigenspectra, 
and the resulting distribution of estimated risk values for the entire QSO sample.
Despite the fact that most of the variance
is contained within the first few eigenspectra, dozens - up to 200 - eigenspectra are needed to 
preserve all the information in the spectra.  For the purposes of this study, reducing the noise and
identifying the outliers, that is not a problem.
The risk values are scaled such that they are similar
to the $\chi^2/\nu$ statistic;
a value of unity indicates a fit that is consistent with the pixel-to-pixel uncertainties.

\subsubsection{Basing the Reconstruction on Spectral Signal/Noise}

While estimated risk provides the optimum number of eigenspectra, $N_{opt},$ required to fit any given
QSO spectrum, ``perfect'' reconstruction of {\it all} QSOs in the sample compromises our second
goal of finding rare or unusual objects --- or even objects that should not have been included
in the sample in the first place.
We will thus not use the $N_{opt}$ values directly, but to inform instead a {\it general} criterion
for fitting the spectra based solely on their spectral signal/noise (S/N) ratios.

\begin{figure}
\epsscale{0.80}
\plotone{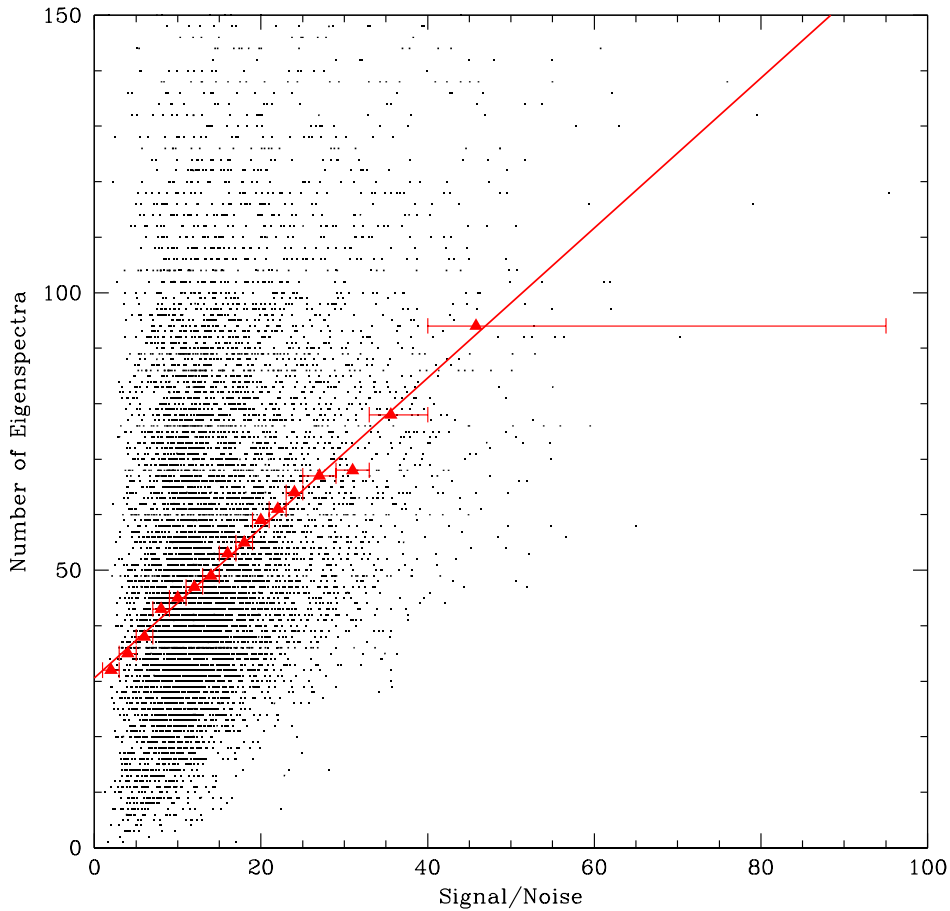}

\caption{The optimal number of eigenspectra needed to minimize the estimated risk is plotted against the 
signal-to-noise ratio for the 9800 individual QSO spectra.  The red triangles show the location of
median number of eigenspectra needed
for subsets of the sample ordered by signal-to-noise.  The error bars show the S/N
extent of each subset.  The red line shows a linear fit to the median points.}
\label{fig:sn}
\end{figure}

Use of S/N as a guideline for understanding how many eigenspectra to fit to a noisy QSO spectrum
makes objective sense.  Noise introduces a source of variance to the spectra,
at some point dominating the information contained in the higher eigenspectra,
which on average capture increasingly small portions of the total variance.
The number of eigenspectra needed thus ought to be related to the S/N
of the spectra, with higher S/N justifying the use of more eigenspectra.
Figure \ref{fig:sn} shows the relationship between
$N_{opt}$ calculated as described above, and the S/N
of each spectrum.  The red points show medians of subsets of the sample; these are very well
approximated by a linear relationship:

\begin{equation}
N_{opt} = 30.54 + 1.35(S/N)
\label{eqn:nopt}
\end{equation}

If we use this criterion to determine the number of eigenspectra needed to reconstruct
the QSO spectra instead of that provided directly by estimated risk, we find that
the average risk value for the sample only goes from 0.858 to
0.873, less than a 2\%\ increase. 
The distribution of points in Figure \ref{fig:sn} appears
to show a wide range in the ``true'' $N_{opt}$ as provided by estimated
risk about this line, but since the most of the variance is provided by the lowest
eigenspectra (as is shown in Figure \ref{fig:var}), this in practice means the two different criteria
for $N_{opt}$ will produce only modest differences in the
quality of the representations.

In passing, we also emphasize a highly practical motivation for using
$N_{opt}$ as provided by equation (\ref{eqn:nopt}).
Calculation of estimated risk for the
complete QSO sample is an intensive operation that required
more than 500 hours of CPU time.  For each QSO spectrum, we performed
25 reconstructions for each trial $N_{opt}$ value, which itself
could range from 1 to 200 (although we truncated the calculation
early, once it was clear that the risk starting increasing with the
number of eigenspectra).  Each reconstruction required the inversion
of a matrix (up to $200\times 200$).  Given the $\sim 10^4$ spectra in the sample,
we were required to perform $\sim 10^8$ matrix solutions.  Evaluating the
S/N for any QSO spectrum, in contrast, is trivial.  For future work using
eigenspectra reconstruction, this suggests an approach in which one
uses only a small subset of the spectra to calibrate a S/N-based
relationship via estimated-risk calculations, such as equation (\ref{eqn:nopt}),
which is then used instead for the complete sample.

\subsubsection{Eigenspectra Reconstruction of the QSO Spectra}

\begin{figure}
\epsscale{0.80}
\plotone{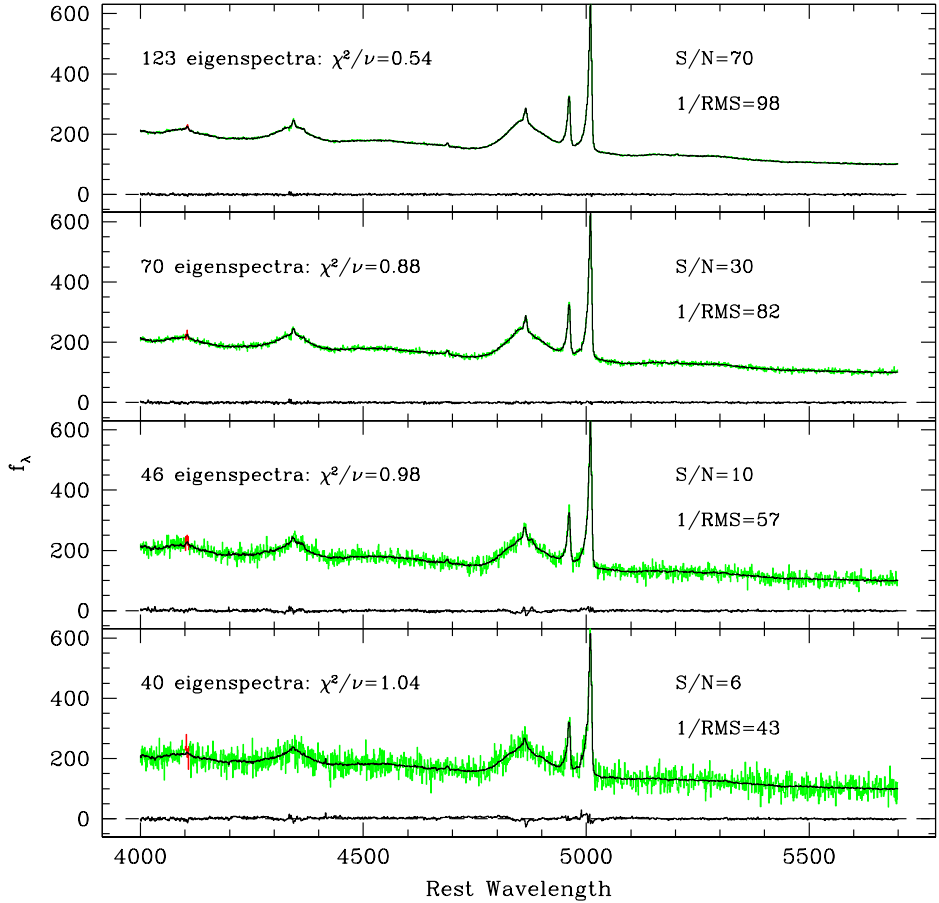}

\caption{The effect of noise on the reconstruction using eigenspectra.  The quasar 
shown is SDSS J105151.44-005117.6.  Input data is shown in green where it is flagged 
as good, in red where it is flagged as bad.  The black line through the data is the 
reconstructed fit.  The black line near the bottom of each panel is the difference between 
the fit in that panel and the original spectrum.  The top panel shows the original spectrum. 
Increasing amounts of Gaussian-distributed noise is added in the lower panels.  Listed 
in each panel is the number of eigenspectra used for optimal reconstruction, the 
$\chi^2/\nu$ value for the fit, the input signal-to-noise of the spectrum, and the effective 
output signal-to-noise.
Note that the [O III] line profiles are recovered with high fidelity,
even in the lowest-S/N test.}
\label{fig:ex1}
\end{figure}

\begin{figure}
\epsscale{0.80}
\plotone{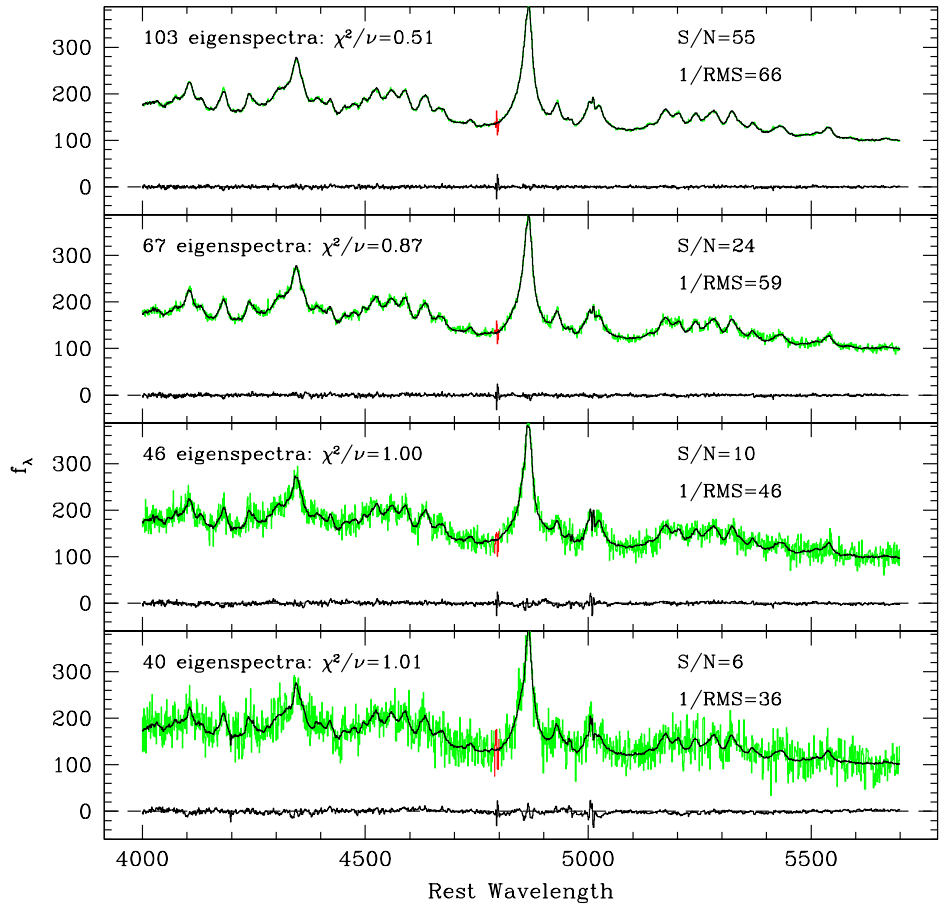}

\caption{As Figure \ref{fig:ex1}, but for the object SDSS J170231.06+324719.6.
Note that the subtle intensity variations in both the red and blue Fe II
complexes are still recovered in even the lowest-S/N test.}
\label{fig:ex2}
\end{figure}

Figures \ref{fig:ex1} and \ref{fig:ex2} demonstrate the effectiveness
of using the eigenspectra reconstructions to
improve the signal-to-noise ratio of spectra.
For each of these two figures, a high signal-to-noise 
spectrum was degraded by adding normally distributed noise.  We chose to start with two 
spectra with very different appearances to show the range of spectral characteristics covered 
by the single set of eigenspectra.  For each degraded spectrum, the new signal-to-noise 
ratio determined how many eigenspectra to use.  The reconstructed spectra are shown as 
black lines through the input spectra.  At the bottom of each panel the difference between 
the reconstruction and the true (high signal-to-noise) spectrum are plotted.  The panels list 
the measured signal-to-noise ratio, the number of eigenspectra used, the resulting value 
of $\chi^2/\nu$, and the reciprocal of the root mean square (RMS) departure of the 
reconstruction to the true 
spectrum.  This last quantity is an effective signal-to-noise for the reconstruction; i.e., it is 
a measure of how well the reconstruction from the degraded low signal-to-noise spectrum 
reproduces the original, high signal-to-noise spectrum.

It is clear that this reconstruction technique 
provides a S/N boost of a factor of 5-7 at an input signal-to-noise ratio of six.
The residuals show that the reconstructions provide
a highly un-biased representation of the spectra, even in the presence
of large amounts of noise.  The residuals have very little low spatial-frequency power;
note that even in the lowest S/N example shown in Figure \ref{fig:ex1}, there are
essentially no errors in the reconstructed widths and intensities of the lines, both 
broad and narrow.
Looking at the other example in Figure \ref{fig:ex2},
it is especially impressive that the subtle ``corrugations'' in both the red and blue
Fe II complexes are recovered in the lowest S/N case.
In contrast, many general linear filters that might be applied to reduce the noise
would certainly broaden such features in both QSOs.
In short, the eigenspectra reconstructions do extremely well at recovering the essential
features in even very low-S/N spectra.
\footnote{Note
that the $\chi^2/\nu$ values for the high signal-to-noise spectra are low because their noise is
included in the eigenspectra.}

\begin{figure}
\epsscale{0.90}
\plotone{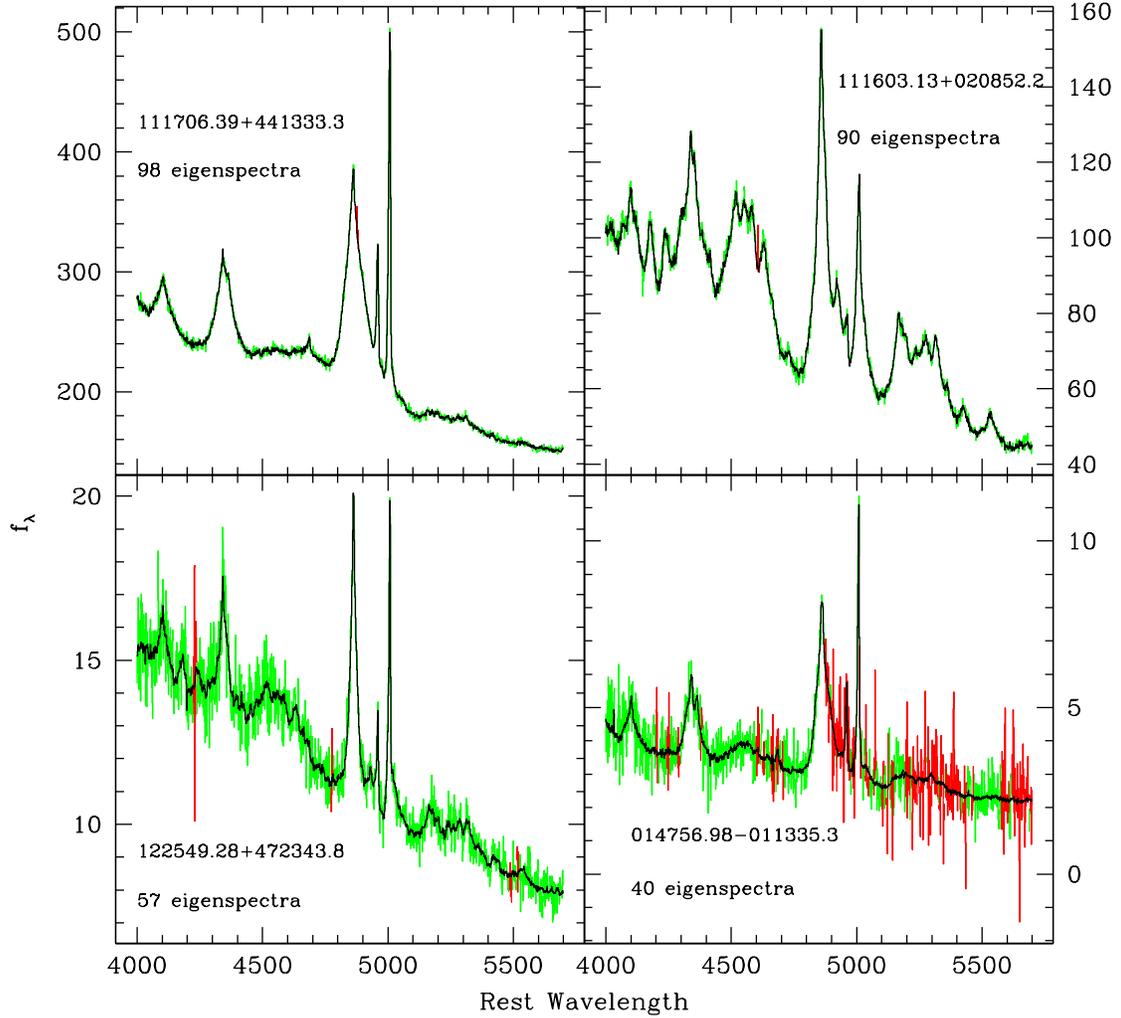}

\caption{Examples of reconstructions using the eigenspectra.  The original input data 
are shown in green (flagged as good) and red (flagged as bad).  In each case, the 
optimum number of eigenspectra to be used for reconstruction is derived from the 
average signal-to-noise of the spectrum.  Higher signal-to-noise input spectra warrant 
the inclusion of more eigenspectra in their reconstruction.}
\label{fig:ex3}
\end{figure}

After establishing the success of this technique with the high signal-to-noise subset that 
had been used to generate the eigenspectra, the entire sample of 9,800 low-redshift 
quasars was reconstructed in this way.  Examples of spectra with a range of 
signal-to-noise and spectral characteristics are shown in Figure \ref{fig:ex3}.  Original data is 
shown in green, except for regions flagged in the original data as bad, typically because 
of poor sky subtraction, which are plotted in red.

The weights with 
which the input spectra can be reconstructed are available in tabular form electronically.  
Table \ref{tab:spec} gives the format of that data.  Thus, this electronic table represents our best attempt at 
reconstructing the spectra of all the low-redshift QSOs in the SDSSQ4 catalog.

\section{Properties of the Sample in the Eigenspace}

\begin{figure}
\epsscale{0.90}
\plotone{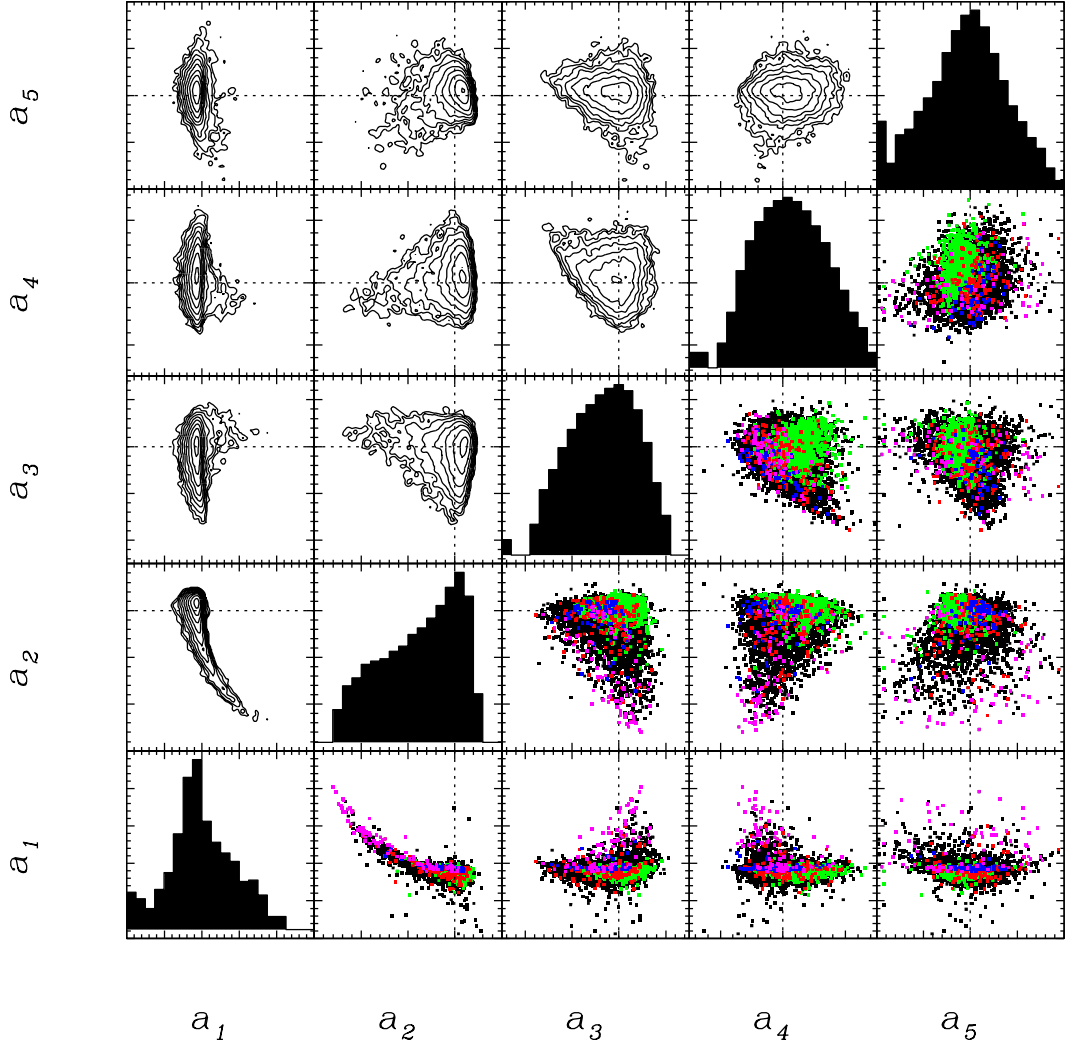}

\caption{Panels show the distribution of weighting factors, $a_i$, for the entire sample of 
9,800 objects on eigenvectors 1 through 5.  Dotted lines indicate the origin in each panel.  
The panels in the upper left show the density distributions as contours.  The panels in the 
lower right show four subsets superposed on the entire sample (plotted as black dots).  
The green points are identified as NLS1s by \citet{zho06}; the red points 
are identified as radio-loud by \citet{kim08}; the blue points are identified as objects 
having double-peaked Balmer lines by \citet{str03}; and the magenta points are those 
which we select as lacking broad emission lines. 
The panels along the diagonal show the histogram of weighting factors for each 
eigenvector on a logarithmic scale.}
\label{fig:esc}
\end{figure}

The multi-dimensional distribution of the eigenspectra coefficients or weights 
may contain interesting information about the overall properties of the sample, or reveal 
interesting sub-populations of the complete set. Visualization and analysis of the 
distribution of the sample in a hyperspace of large dimension is challenging; however, 
for the present case we are just interested in the gross distribution of the first few 
components, and so simple two-component plots suffice to show the most important details.
Figure \ref{fig:esc} plots the distribution of weights for eigenspectra 
1 through 5. In all these plots, and several others showing higher components, the bulk 
of the points always fall within a cloud around (0,0), shown by dotted lines in each panel 
of Figure \ref{fig:esc}, with the exception of  E1, which has a non-zero mean.   

Note that, in general, the distributions of weights are not symmetric or uniform, though 
there is no correlation between the weights on any pair of eigenspectra, by construction.  
The points that are obvious distant outliers in the panels in Figure \ref{fig:esc} are frequently objects 
having large gaps in their spectra or spectra of very low signal-to-noise, but the overall 
impression that the clouds of points have extensions that might be characterized as ÒtailsÓ 
or ÒfansÓ is accurate.  We were curious whether these extensions might represent subsets 
of particular interest for studies that focus on a particular type of object, Narrow-line 
Seyfert 1s (NLS1), for example.  Thus, we cross-matched the SDSSQ4 sample with 3 lists 
from such studies: the \citet{zho06} list of 2000 NLS1s drawn from SDSS Data Release 3, 
the \citet{kim08} list of 1300 QSOs detected by FIRST, NVSS, WENSS, and SDSS, and 
the \citet{str03} list of 116 AGN from an early SDSS release with double-peaked Balmer 
line profiles.  Each of these lists includes some objects not contained within our overall sample, 
because of the redshift limit or luminosity limit or the allowance for non-stellar objects, and 
so the actual number cross-matched in each case is somewhat less than the full lists.  To 
these three lists we added a fourth subset: those objects identified in section 4 below as 
having narrow emission lines only in this spectral region.  The objects cross-matched with 
each of these lists are shown as colored points in the lower right panels of Figure \ref{fig:esc}.
It can be seen that these colored points are somewhat segregated; in most of these planes, 
NLS1s do not strongly overlap with radio-loud QSOs.  Our intent is not to derive a formula 
or procedure for finding a particular type of object, but to point out that a rational starting 
point for such a procedure would be to examine the regions in which known objects of a 
particular type cluster.  We note that this technique was used by \citet{str03} to enhance 
the density of candidate double-peaked profile objects within the sample they searched 
by a factor of five.

\begin{figure}
\plotone{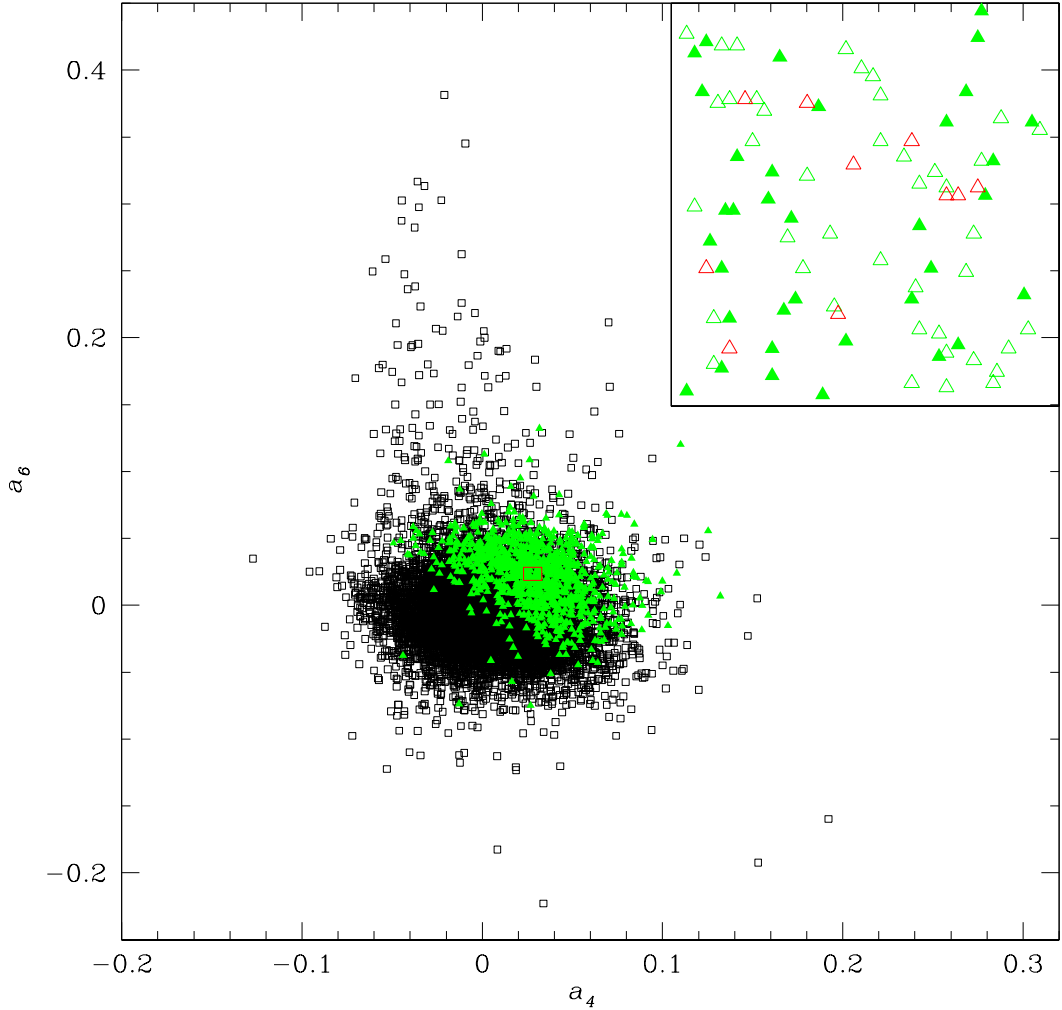}

\caption{The entire sample (open black squares) in the e4-e6 plane, with objects classified
as NLS1s by \citet{zho06} plotted as solid green triangles.
This plane appears to be especially well-suited to selecting NLS1s from the rest of the sample.
The small red rectangle shows the
location of the region expanded in the upper right.  Within that region, previously classified
NLS1s are shown as solid green triangles, new points are plotted as open triangles - green 
if they are NLS1s, red if they are not.  Note that nearly all of the objects within this
box are NLS1s.}
\label{fig:nls1}
\end{figure}

We take this demonstration one step further in the case of NLS1s, for which we have 
enough known points to define a preferred region clearly.   Inspection of the histograms 
of weights for the NLS1s and comparison with those for the entire sample shows that E4 and E6
(Figure \ref{fig:nls1}) provide the greatest discrimination in separating NLS1s from the rest of 
the objects.  We define a small preferred region for NLS1s that is centered on the mean 
values for their weights on E4 and E6.  Within this region there are 84 objects, of which 
32 are already classified as NLS1s.  Of the 52 new objects, 42 have
H$\beta$ FWHM values below 2200 km s$^{-1}$, the threshold for classification as a 
NLS1.   Thus, the fraction of NLS1s in this region is 85-90\% 
of all objects.  This can be compared with a fraction of about 8.5\% for the entire quasar 
eigenspace, obtained by dividing the number of NLS1s from the \citet{zho06} list that 
are successfully cross-matched with our quasar sample (1097) divided by the number 
of SDSS spectra of objects classified as quasars (12,824) that they searched. 

\subsection{A Search for Clusters in the Eigenspace}

In all two-component projections of eigenspace that we examined,
the parameters form a single aggregate;
but out of concern that these plots were too crude to uncover multiple clusters
in the much higher-dimensioned eigenspace, we attempted a simple search
for such features based on binning the QSOs in a 5-dimensional space defined by their E1-E5 
coefficients.  Since these are the most important coefficients, we would expect any 
clustering to be manifest in this sub-space.

The procedure was to bin the QSOs in the 5-space, selecting bins that span the 
interesting range of the coefficients, are small enough to dissect the central cloud of 
points seen in the two component plots, but are not so small as to make the problem 
difficult. Having then binned the sample, we performed a simple analysis to see if various 
combinations of the bins defined separate aggregates.  We first applied a variable threshold 
for the binned-density of the QSOs, and then attempted to link bins that rose above the 
threshold into connected aggregates.  Neighboring bins were linked when they differed 
from each other by a $\pm$1 increment in one and only one index in the 5-space, and 
thus had the highest possible connectivity (the analogy in 3-space would be to link cubes 
with faces in common, but not those that only shared an edge or corner).  For all threshold 
choices we were always able to link the surviving bins into one large aggregate (with a 
modest number of bins remaining isolated).  We could find no evidence for more than 
one cluster in the space that we examined; there is no evidence for bi- (or higher order) 
modality based on the characteristics within this spectral region.

\section{The Outliers}

One of our goals in developing these tools is to identify those objects in the sample that are unique or
unusual, either because they are not appropriately part of the sample, or because they
are individually interesting objects.  Finding these objects requires a two-step process: first,
use of the mathematical tools to produce a subsample of candidates, and second, visual
inspection and classification of the subsample.  Thus, it is a combination of mathematical
and astronomical knowledge that is required.  

For the first step, we tried three algorithms for identifying outliers or contaminants:

\begin{itemize}

\item Most of our candidate outliers came from those spectra which have a large risk value (or $\chi^2/\nu$)
when reconstructed using the number of eigenspectra given by our linear signal-to-noise formula. We
adopted a value of 1.50 for this threshold, but we stress that we found no value that clearly distinguished
outliers.  The adopted value created a subsample that was not onerous to inspect visually.
This technique isolates both those objects whose spectra are not well fit and those that require
significantly more eigenspectra before they were well fit.

\item  We examined the 50 objects that were farthest from the origin of the eigenspace,
using the first seven coefficients to compute a distance.

\item  We identified the 50 objects that were the most ``isolated'' in eigenspace, that is,
those that had the largest distances (based on the first seven coefficients) from their
closest neighboring objects (as might be expected, this set had some overlap with objects
identified by their distance from the origin).

\end{itemize}

The total number of candidate outliers selected by these three methods is 250 objects.
Of the 100 objects identified by isolation or distance from the origin, 70
are not in the sample defined by the risk value alone, and 30 are selected by both
poor fit and by one of the eigenspace criteria.
None of these tests show a sharp distinction between what will be a
contaminant and what will not.  Any particular extreme characteristic may signify an object that
is not a quasar or it may just be an object with rarely seen features.  Additionally, some rarely seen
features have important physical implications and some do not.  And conversely, some contaminants
are fit tolerably well despite their unusual features.  Thus, there is no rigorous criterion for identifying 
objects that are unambiguously outliers, only procedures for producing subsets that may then
be examined singly.

We visually examined the spectra of these 250 objects, and we classified them into the following 
seven categories (number of objects): no broad lines (48), broad lines present but very weak (34), 
anomalous [O III] line profiles (21), anomalous H$\beta$ profile (9),  other miscellaneous
peculiarity in spectrum (3), bad region not masked properly (30), and no obvious specific problem (124).  
The last category includes objects that lie near the extrema of the characteristics represented by
the high signal-to-noise objects that were used to construct the eigenspectra -- for example,
strong Fe II emission, narrow broad lines, or relatively strong narrow [O III] $\lambda$4363 emission.
It also includes objects with strong galactic Na I D absorption.
Note that the sum of the numbers above exceed 250 because 19 of the objects are counted in two
categories, 16 with no broad lines and
anomalous [O III] line profiles, and 3 with weak broad lines and anomalous [O III] profiles.  
Below we briefly describe the objects in these different categories.  The objects in the first five 
categories listed above are identified in Table \ref{tab:out}.

\begin{figure}
\epsscale{0.90}
\plotone{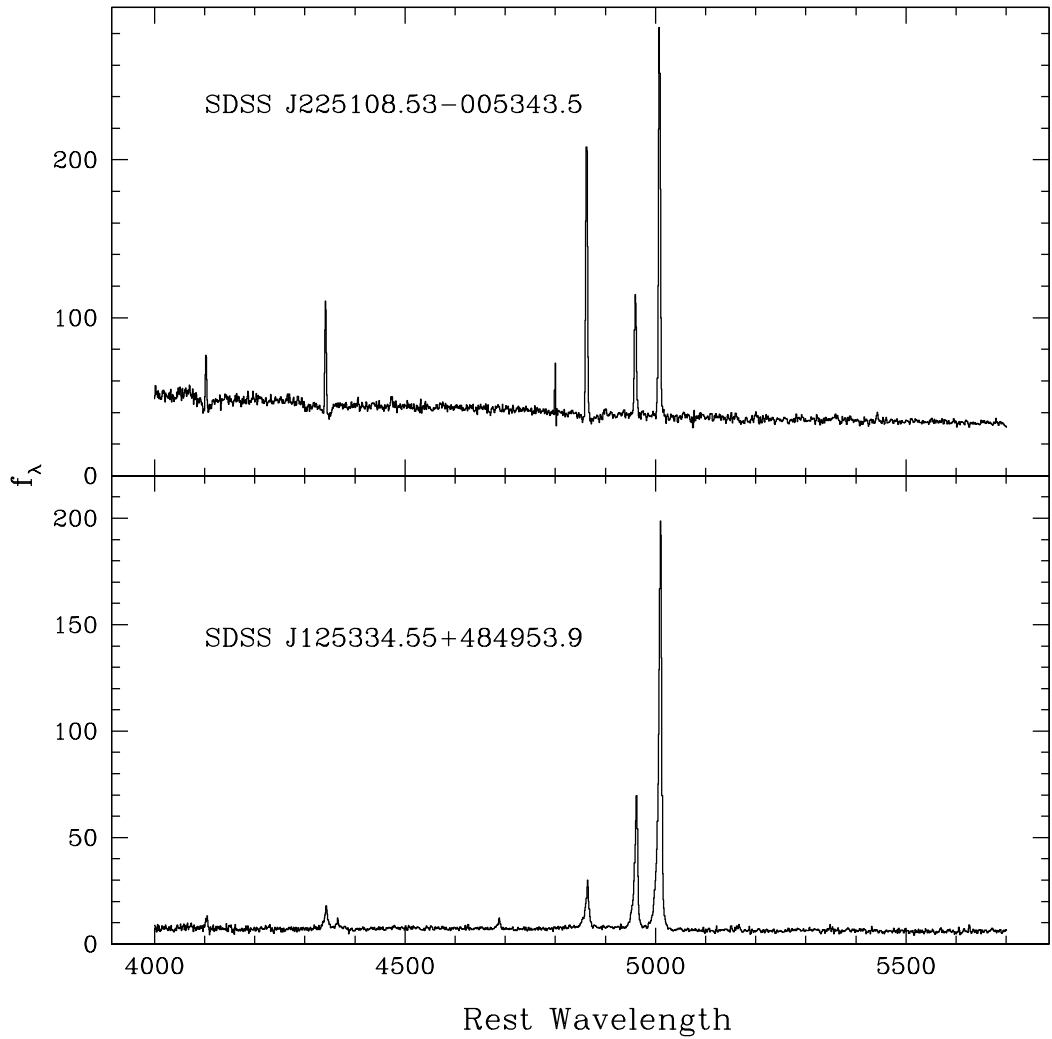}

\caption{Examples of two objects with no broad lines.  The top panel shows the spectrum of
SDSS J225108.53-005343.5, a high luminosity star-forming galaxy.  The lower panel show
the spectrum of SDSS J125334.55+484953.9, a type II quasar.}
\label{fig:nobl}

\end{figure}

The objects with no broad lines should not have been in the SDSS QSO catalog,
which has  a requirement that at least one emission line must have a width greater than 1000
km s$^{-1}$.  These objects fall into two categories: those that are type II quasars and those that are 
galaxies with strong emission lines photoionized by hot stars.  Figure \ref{fig:nobl} shows a typical example of 
each of these.  Among the type II quasars, a small fraction show broad H$\alpha$ or broad 
Mg II $\lambda$2800, two lines that were not included in the rest wavelength range that we 
analyzed.

\begin{figure}
\epsscale{0.90}
\plotone{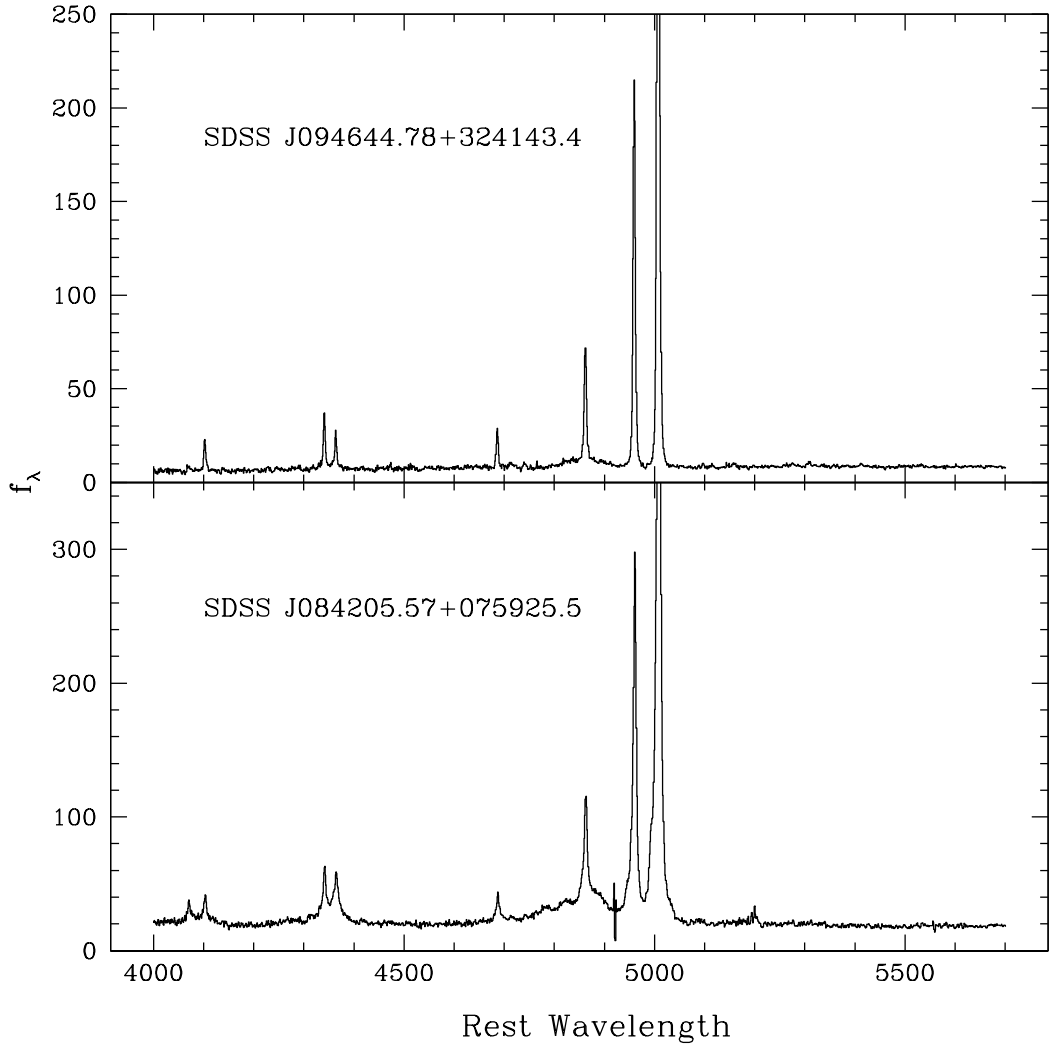}

\caption{Examples of two objects with very weak broad lines.}
\label{fig:wbl}

\end{figure}

A number of objects show quite weak broad H$\beta$.  Although these are legitimately members of the
quasar catalog, they are unusual either in having a very large Balmer decrement in the broad lines,
or relatively weak broad line emission overall.   Figure \ref{fig:wbl} shows two examples of these objects.

\begin{figure}
\epsscale{0.90}
\plotone{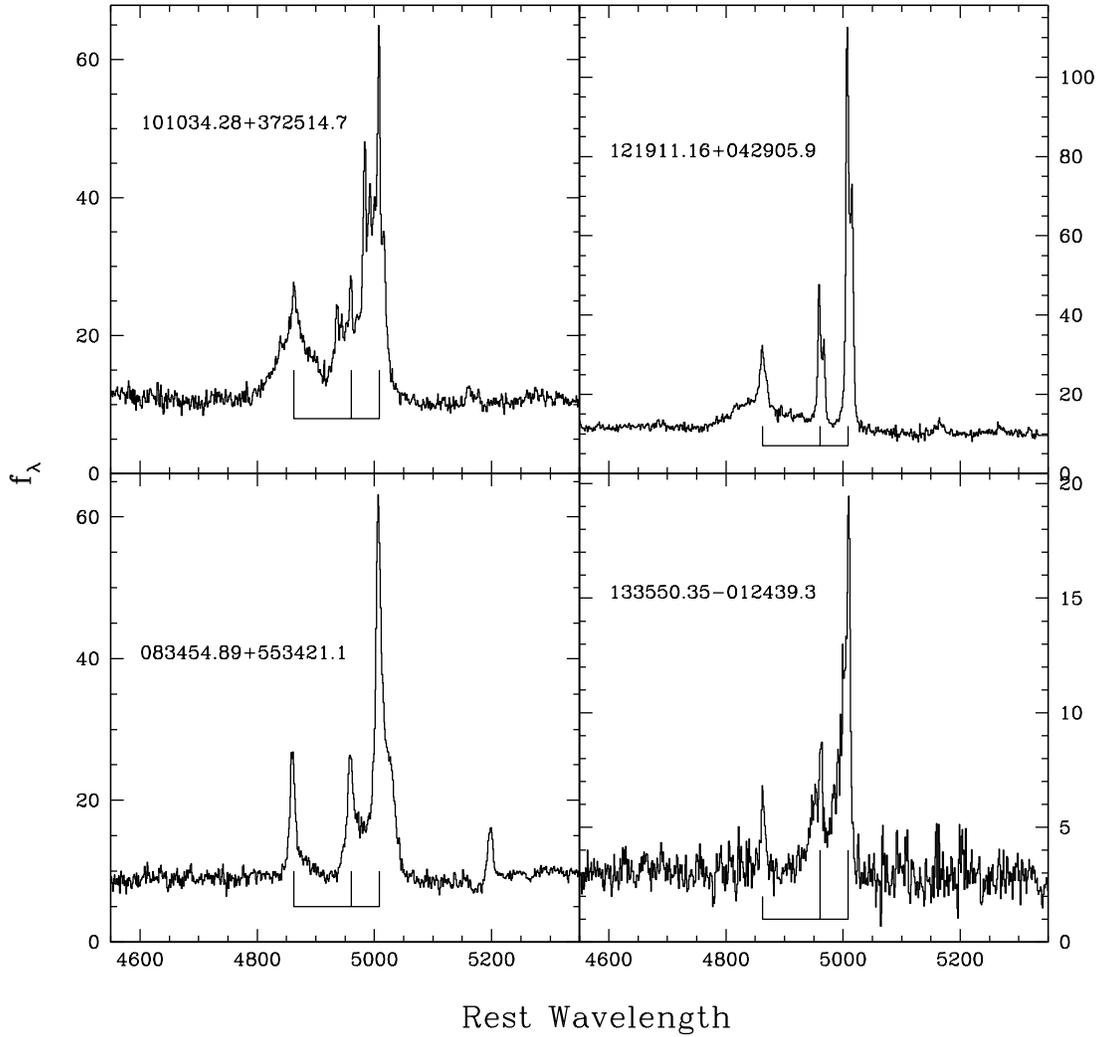}

\caption{Four examples of objects with [O III] $\lambda\lambda$4959, 5007 lines having 
unusual structure or width. Each panel shows the nominal location of H$\beta$ and the 
[O III] $\lambda\lambda$4959, 5007 doublet.  The objects in the top panels have broad 
H$\beta$ emission, while those in the lower panels do not.}
\label{fig:oiii}

\end{figure}

Several examples of objects that show unusual structure in the [O III] lines are shown 
in Figure \ref{fig:oiii}.  Some of these show complex multiple peaks, often shifted to the blue from
the systemic redshift.  These lines are presumably caused by outflows driven by large
scale bursts of star formation \citep{hwa07,lip03,vei99}. We note that 
SDSS J 101034.28+372514.7 is a new object in this class, showing [O III] emission 
with five clear peaks from +440 to -1480 km s$^{-1}$ relative to the systemic velocity of 
the object.  Other objects show very broad or 
double emission lines with splittings of a few hundred km s$^{-1}$.  These may be due to
one or more of four effects:  (1) supernova-driven outflows or winds from massive starbursts 
\citep{hec90, lip03}, (2) biconical outflows from the AGN \citep{cec02}, 
(3) biconical or complex radiation pattern that 
illuminates material with different velocities within the gravitational potential of the 
host galaxy \citep{wil94, pog89}, and (4) binary black hole systems \citep{zho04}.  

\begin{figure}
\epsscale{0.90}
\plotone{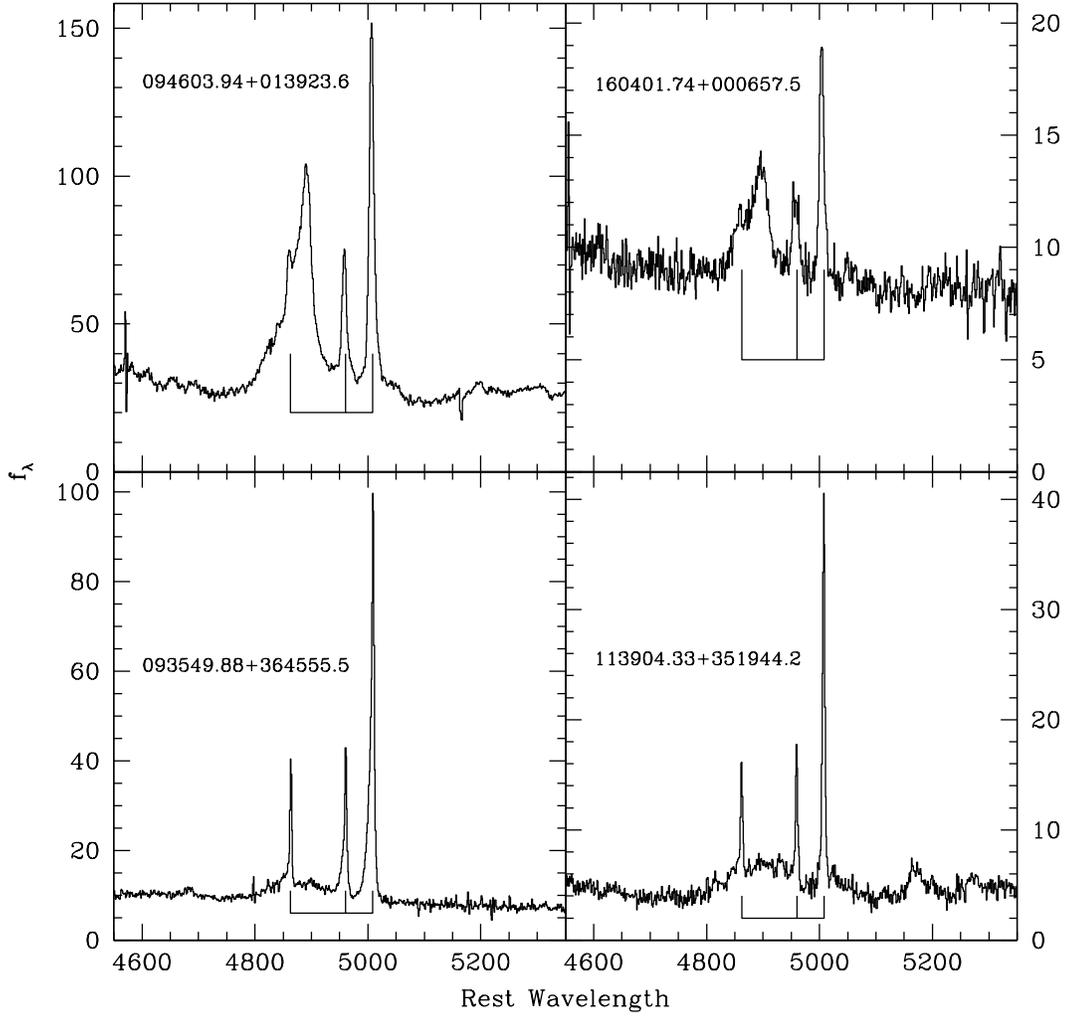}

\caption{Four examples of objects with unusual structure in the H$\beta$ line.   Each 
panel shows the nominal location of H$\beta$ and the [O III] $\lambda\lambda$4959, 5007 
doublet.}
\label{fig:phb}

\end{figure}

Several of the objects that exhibit peculiar  H$\beta$ profiles are shown in Figure \ref{fig:phb}.  
The phenomenon of offset or multiple 
 peaks might be due to (1) non-uniform distribution of emitting material in the broad line 
 region, (2) emission from the accretion disk surrounding the supermassive black hole
 \citep{era94, str03}, 
 or (3) one or more nuclei offset kinematically from the narrow line region, which, 
 presumably, represents the systemic redshift of the host galaxy \citep{bor09}.  
 
\begin{figure}
\epsscale{0.90}
\plotone{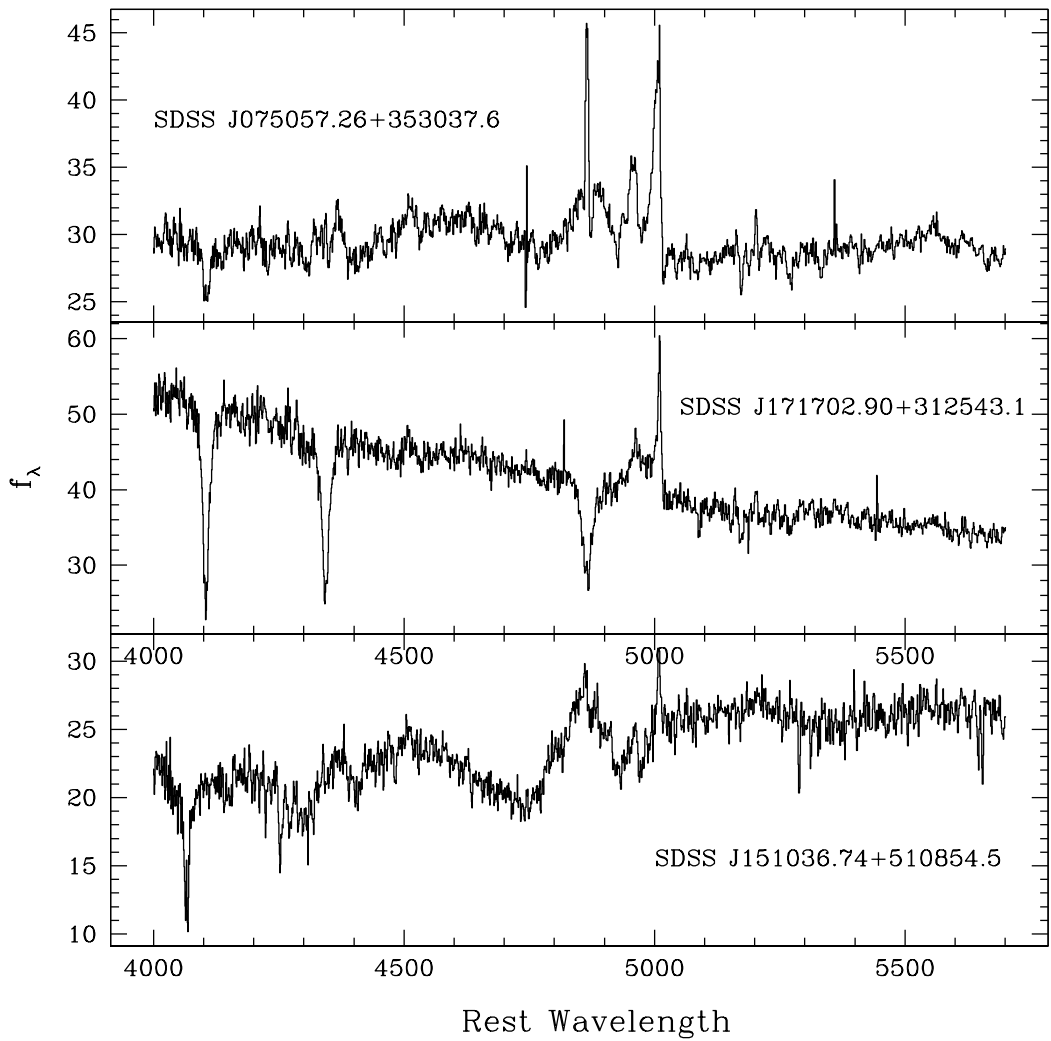}

\caption{The three objects identified as miscellaneous outliers.  The top two panels show
objects with strong Balmer absorption from hot stars in the host galaxy.  The bottom panel
shows an object which is the superposition of a QSO at redshift 0.4494 and a galactic star. }
\label{fig:misc}

\end{figure}

 The remaining objects in our list of outliers comprise those that do not fall into the 
 more common categories.  The spectra of these three objects are shown in Figure
 \ref{fig:misc}.  We describe their characteristics below:

SDSS J075057.26+353037.6 (z = 0.1759) and SDSS J171702.90+312543.1 (z=0.1577) 
are AGNs in galaxies with strong early-type absorption  spectra.  This produces the 
appearance of an absorption feature in the broad H$\beta$ emission line.

SDSS J151036.74+510854.5 (z = 0.4494) has a spectrum that appears to be a 
 superposition of a late type galactic star and a typical quasar at the listed redshift.  

This project was begun when the DR5 version of the archive was the most recent
release.  During 
the work, the DR7 version was released, and so we applied this analysis to the
5500 additional objects identified as quasars 
with z $<$ 0.7 in the SDSS DR7 archive.  
 We note that this sample has not been vetted in the same way as the DR5 quasar 
 catalog, but consists of everything that the SDSS spectroscopic reduction pipeline 
 classified as a quasar.  For this sample, 160 outliers were identified.  These fall 
 into the same categories as our principal sample, with one exception.  The object 
 SDSS J153636.22+044127.0 (z = 0.3889) is a quasar with two, very clear, 
 broad-line redshift systems, separated by about 3500 km s$^{-1}$.  This is, to our 
 knowledge, a unique object (though we have identified objects with much less
 prominent multiple peaks above), and it is a candidate for a sub-parsec supermassive 
 binary black hole system.  A complete analysis of the spectrum and discussion 
 of interpretations is given in \citet{bor09} and \citet{lau09}. Additional observations 
 and interpretations have been published \citep{wro09, dec09, tan09, cho10}, but
 the nature of this system remains ambiguous. 
 
\section{Summary and Conclusions}

We have demonstrated the usefulness of the Karhunen-Lo\`{e}ve transform for 
simplifying the information in large samples of QSO spectra.  We have developed 
procedures for applying the KL transform to a restricted wavelength region in the 
spectra of 9,800 QSOs from the Sloan Digital Sky Survey, adopting 
a concept described by \citet{con99}.

Rather than attempt to interpret the eigenspectra in terms of physical properties or 
processes, we take advantage of the mathematical aspects of the KL transform. 
Because the eigenvectors are ordered to account for decreasing fractions of the 
sample-wide variance, reconstruction of a spectrum using a limited number of 
eigenvectors can fill in gaps in the data or improve the S/N, provided 
that the spectrum represents a valid member of the ensemble used to create the 
eigenspectra.  We have developed a S/N-based rationale for selecting the number of 
eigenvectors to be used for any given spectrum that results in an increase in 
signal-to-noise by as much as a factor of 6 in certain cases.  The number of 
eigenspectra used ranged from 30 to 200 for our sample, as the S/N of 
the input spectra ranged from 1.3 to 95.  The improved S/N in the resulting set 
allows for better measurement and analysis of the properties of low-z SDSS QSOs.

The eigenspace is not uniformly populated, and the plots of spectrum weights 
on pairs of eigenspectra show tails and plumes that represent the frequency of 
various spectral characteristics.  Despite this lack of symmetry, we see no 
evidence for more than one cluster, which would have indicated one or more 
discontinuous subclasses of objects.  Using the locations of the 9,800 QSO spectra in our 
eigenspace, we show that subclasses of QSOs, e.g., NLS1s, are concentrated 
in certain regions.  This allows an improvement in efficiency in searches for such 
objects.

We have explored several ways of identifying objects that are unique or rare.  
Such objects are of interest either because they are not valid members of the 
sample and should be removed, or because they are extreme examples of 
important phenomena that may provide insight into our understanding of the 
physics.  Such objects may be selected for having extreme values of one or 
more eigenvector weights, for being in isolated regions of the eigenspace, or 
for having $\chi^2$ or risk values that indicate a poor fit.  

Adopting a risk value of 1.5 as the threshold indicating an unsatisfactory fit, and
adding objects that are far from nearest neighbors or far from the origin of the
eigenspace,
we identified 250 objects for visual inspection.  Removing those objects with poor data, 
we are left with 126 objects that we classify into several categories.  Some of these
-- objects with no broad emission lines -- should not have been included in the 
quasar catalog.  A number are physically interesting objects, including objects 
with complex or offset emission line profiles, and one object,
 SDSS J153636.22+044127.0, which may be a merging black hole binary system.

\acknowledgments

We thank Andy Connolly for a helpful conversation during the time that we were 
developing the software.  We thank an anonymous referee
who pointed us towards the technique of cross-validation, and Joey Richards for
helping us through the implementation and interpretation of this technique.   
This work utilized several 
software packages that are made available free, and we wish to thank the individuals 
or groups that produced them and made them available.  The software that we 
developed to find the eigenvectors and reconstruct the spectra is based on a 
program (pca\_public.f) written and distributed by Paul Francis.  The initial processing 
of the SDSS FITS spectra was done using IRAF, which is distributed by the National 
Optical Astronomy Observatory (NOAO).  Much of the exploration of the eigenspace 
was done using TOPCAT, currently developed and supported by the Astrogrid project 
in the UK.  Finally, much of the visual inspection of the spectra and their 
reconstructions, and the production of the final figures used SM, written by Robert 
Lupton and Patricia Monger.  NOAO is operated by the Association of Universities 
for Research in Astronomy (AURA) under cooperative agreement with the National 
Science Foundation. This research has made use of the NASA/IPAC Extragalactic 
Database (NED) which is operated by the Jet Propulsion Laboratory, California 
Institute of Technology, under contract with the National Aeronautics and Space 
Administration.

This paper has made use of the data from the SDSS. Funding for the SDSS and 
SDSS-II has been provided by the Alfred P. Sloan Foundation, the Participating 
Institutions, the National Science Foundation, the U.S. Department of Energy, the 
National Aeronautics and Space Administration, the Japanese Monbukagakusho, 
the Max Planck Society, and the Higher Education Funding Council for England. 
The SDSS Web Site is \url{http://www.sdss.org/}.

    The SDSS is managed by the Astrophysical Research Consortium for the 
Participating Institutions. The Participating Institutions are the American Museum 
of Natural History, Astrophysical Institute Potsdam, University of Basel, University 
of Cambridge, Case Western Reserve University, University of Chicago, Drexel 
University, Fermilab, the Institute for Advanced Study, the Japan Participation Group, 
Johns Hopkins University, the Joint Institute for Nuclear Astrophysics, the Kavli 
Institute for Particle Astrophysics and Cosmology, the Korean Scientist Group, the 
Chinese Academy of Sciences (LAMOST), Los Alamos National Laboratory, the 
Max-Planck-Institute for Astronomy (MPIA), the Max-Planck-Institute for Astrophysics 
(MPA), New Mexico State University, Ohio State University, University of Pittsburgh, 
University of Portsmouth, Princeton University, the United States Naval Observatory, 
and the University of Washington.

 {\it Facilities:} \facility{Sloan}.

\clearpage

\clearpage

\clearpage

\clearpage

\clearpage

\clearpage

\clearpage

\clearpage

\clearpage

\clearpage

\clearpage

\clearpage

\clearpage

\begin{deluxetable}{ccccc}

\tablecaption{Eigenspectra}
\tablewidth{0pt}
\tablenum{1}

\tablehead{\colhead{Column} & \colhead{Format} & \colhead{Units} & \colhead{Label} & \colhead{Description} \\ 
\colhead{} & \colhead{} & \colhead{} & \colhead{} & \colhead{} } 

\startdata
1 & f7.2 & Angstroms & Lambda & Wavelength of center of pixel \\
2 & f10.6 & \nodata & E1 & Value of eigenspectrum 1 pixel \\
3 & f10.6 & \nodata & E2 & Value of eigenspectrum 2 pixel \\
4 & f10.6 & \nodata & E3 & Value of eigenspectrum 3 pixel \\
 &  &  &  &  \\
201 & f10.6 & \nodata & E200 & Value of eigenspectrum 200 pixel \\
\enddata

\tablecomments{Table available electronically.}
\label{tab:es}
\end{deluxetable}
\begin{deluxetable}{cccc}

\tablecaption{Reconstruction Weights for the QSO Sample}
\tablewidth{0pt}
\tablenum{2}

\tablehead{\colhead{Column} & \colhead{Format} & \colhead{Label} & \colhead{Description} \\ 
\colhead{} & \colhead{} & \colhead{} & \colhead{} } 

\startdata
1 & a18 & SDSS Name & SDSS DR5 Object Designation \\
2 & f6.4 & z & Redshift from SDSSQ4 catalog \\
3 & f8.2 & Norm & Normalization to units of 10$^{-17}$ erg cm$^{-2}$ s$^{-1}$ \AA$^{-1}$ \\
4 & i2 & N & Number of eigenspectra to be used \\
5 & f5.3 & $\chi^2/\nu$ & Cross-validation value (reduced chi-squared) \\ & & & of reconstructed spectrum \\
6 & f7.4 & a1 & Weight for eigenspectrum 1 \\
7 & f7.4 & a2 & Weight for eigenspectrum 2 \\
8 & f7.4 & a3 & Weight for eigenspectrum 3 \\
 &  &  &  \\
205 & f7.4 & a200 & Weight for eigenspectrum 200 \\
 \\
\enddata

\tablecomments{Only weights for eigenspectra up to N are given.  Table available electronically.}
 \label{tab:spec}
\end{deluxetable}

\begin{deluxetable}{lll}
\tablecaption{Outliers}
\tablewidth{0pt}
\tablenum{3}
\tablehead{\multicolumn{3}{c}{SDSS J}}
\startdata
\multicolumn{3}{c}{Objects with no broad lines} \\
\hline

005621.72+003235.6 & 103951.49+643004.1 & 141718.61+341709.5 \\
010750.47-005352.9 & 104014.42+474554.7 & 142047.97+032557.2 \\
013416.34+001413.5 & 110550.54+112702.1 & 142209.95+250927.7 \\
020038.67-005954.5 & 110927.02+641833.3 & 142939.80+395935.3 \\
020655.71+010826.6 & 113023.80+115342.4 & 150407.51-024816.5 \\
022250.39-074709.4 & 115314.35+032658.5 & 152238.10+333135.8 \\
030822.35-001114.7 & 115718.34+600345.6 & 153246.68+495458.4 \\
032029.78+003153.6 & 121522.77+414621.0 & 155059.37+395029.5 \\
040144.54-060538.7 & 122845.74+005018.7 & 160641.42+272556.9 \\
081352.97+381602.5 & 123006.79+394319.3 & 163411.91+231348.1 \\
083454.89+553421.1 & 123309.87+154952.2 & 163653.37+245746.4 \\
085946.67+010812.6 & 125334.55+484953.9 & 171600.87+274414.3 \\
090226.89+300607.0 & 132958.47+025623.0 & 171642.60+311006.9 \\
091345.48+405628.2 & 133550.35-012439.3 & 224027.05+004347.4 \\
092152.45+515348.1 & 134559.16+414918.4 & 225108.53-005343.5 \\
094311.57+345615.8 & 135128.15-001016.9 & 230321.73+011056.4 \\
\hline
\multicolumn{3}{c}{Objects with weak broad H$\beta$}\\
\hline
033408.38-004235.6 & 101557.61+483759.6 & 143241.11+430039.5 \\
073422.20+472918.8 & 102618.97+563249.2 & 155049.50+350915.7 \\
080413.87+470442.8 & 105635.05+414602.5 & 155059.89+031559.5 \\
082049.66+560659.1 & 110718.10+083442.8 & 160041.30+082309.7 \\
082857.99+074255.7 & 110844.10+462308.9 & 160926.83+083221.6 \\
084205.57+075925.5 & 121118.66+143810.4 & 211343.19-075017.6 \\
092501.78+274607.9 & 131945.96+053002.8 & 220408.40+113603.9 \\
092751.11+343103.6 & 132515.57+050156.4 & 224256.47+005155.2 \\
093728.59+053750.3 & 134303.59+521626.8 & 230004.46-095432.7 \\
094644.78+324143.4 & 140848.33+052355.8 & 230248.88+134553.4 \\
095906.59+510325.3 & 141956.65+060626.8 & 231645.07-001129.4 \\
\hline
\multicolumn{3}{c}{Objects with anomalous [O III] profiles}\\
\hline
005621.72+003235.6 & 094311.57+345615.8 & 122845.74+005018.7 \\
010750.47-005352.9 & 101034.28+372514.7 & 133550.35-012439.3 \\
032029.78+003153.6 & 103951.49+643004.1 & 134559.16+414918.4 \\
082857.99+074255.7 & 104014.42+474554.7 & 135128.15-001016.9 \\
083454.89+553421.1 & 113023.80+115342.4 & 142939.80+395935.3 \\
090226.89+300607.0 & 115718.34+600345.6 & 160641.42+272556.9 \\
092501.78+274607.9 & 121911.16+042905.9 & 224256.47+005155.2 \\
\hline
\multicolumn{3}{c}{Objects with anomalous H$\beta$ profiles}\\
\hline
031719.03-081702.9 & 094603.94+013923.6 & 160401.74+000657.5 \\
083223.36+555850.4 & 113238.41+011811.6 & 212123.90-081310.8 \\
093549.88+364555.5 & 113904.33+465651.1 & 230216.84+135723.5 \\
\hline
\multicolumn{3}{c}{Objects with miscellaneous anomalies}\\
\hline
075057.26+353037.6 & 151036.74+51054.5 & 171702.90+312543.1\\
\enddata

\label{tab:out}
\end{deluxetable}

\end{document}